\documentclass[preprint,12pt]{elsarticle}

\usepackage[utf8]{inputenc}
\usepackage[T1]{fontenc}
\usepackage{amsmath,amsfonts,amssymb}
\usepackage{graphicx}
\usepackage{hyperref}
\usepackage{enumitem}
\usepackage{natbib}
\usepackage{amsthm} 
\newtheorem{theorem}{Theorem}
\usepackage{algorithm}
\usepackage{algpseudocode}
\usepackage{booktabs}
\usepackage{multirow}
\usepackage{array}
\usepackage{soul}
\usepackage{subcaption} 
\usepackage[dvipsnames]{xcolor}
\usepackage{pifont}

\begin{document}

\begin{frontmatter}
\title{Scalable Hierarchical AI–Blockchain Framework for Real-Time Anomaly Detection in Large-Scale Autonomous Vehicle Networks}
\author[1]{Rathin Chandra Shit}
\ead{rathin088@gmail.com}
\author[2]{Sharmila Subudhi\corref{cor1}}
\ead{sharmilasubudhi@ieee.org}
\affiliation[1]{organization={Dept. of Computer Science \& Engg., International Institute of Information Technology}, city={Bhubaneswar}, postcode={751003}, 
state={Odisha}, country={India}}
\affiliation[2]{organization={Dept. of Computer Science, Maharaja Sriram Chandra Bhanja Deo University}, city={Baripada}, postcode={757003}, state={Odisha}, country={India}}
\cortext[cor1]{Corresponding author}
\begin{abstract}
\textbf{Purpose:} The security of autonomous vehicle networks is facing major challenges, owing to the complexity of sensor integration, real-time performance demands, and distributed communication protocols that expose vast attack surfaces around both individual and network-wide safety. Existing security schemes are unable to provide sub-10 ms (milliseconds) anomaly detection and distributed coordination of large-scale networks of vehicles within an acceptable safety/privacy framework.

\textbf{Method:} This paper introduces a three-tier hybrid security architecture HAVEN (Hierarchical Autonomous Vehicle Enhanced Network), which decouples real-time local threat detection and distributed coordination operations. It incorporates a light ensemble anomaly detection model on the edge (first layer), Byzantine-fault-tolerant federated learning to aggregate
threat intelligence at a regional scale (middle layer), and selected blockchain mechanisms (top layer) to ensure critical security coordination.

\textbf{Result:} Extensive experimentation is done on a real-world autonomous driving dataset. Large-scale simulations with the number of vehicles ranging between 100 and 1000 and different attack types, such as sensor spoofing, jamming, and adversarial model poisoning, are conducted to test the scalability and resiliency of HAVEN. Experimental findings show sub-10 ms detection latency with an accuracy of 94\% and F1-score of 92\% across multimodal sensor data, Byzantine fault tolerance validated with 20\% compromised nodes, and a reduced blockchain storage overhead, guaranteeing sufficient differential privacy.

\textbf{Conclusion:} The proposed framework overcomes the important trade-off between real-time safety obligation and distributed security coordination with novel three-tiered processing. The scalable architecture of HAVEN is shown to provide great improvement in detection accuracy as well as network resilience over other methods.  
\end{abstract}

\end{frontmatter}

\section{Introduction}\label{sec:introduction}

The unprecedented rise of autonomous vehicles (AV) has altered the transport industry by providing unparalleled connectivity, intelligence, and an accessible transportation medium to people. These vehicles process multimodal sensor data collected from LiDAR, cameras, radar, and GPS/IMU sensors networked on a CAN (Controller Area Network) bus, generating terabytes of data in a day that require real-time analysis for safe operation \cite{wang21,khan24,Shit2018a}. However, the automotive systems are marred by state-of-the-art security threats characterized by safety-critical domains. Further, the distributed nature of vehicular networks poses significantly higher computational complexity and inherent difficulties while developing efficient cybersecurity frameworks for detecting wide-array of threats in real-time.

The automotive industry is facing a major security paradox where solutions involving AVs are associated with a fundamental trade-off between real-time system performance and distributed security coordination. Traditional anomaly/intrusion detection systems deployed on AV infrastructural networks are based on centralized architectures having a single point of failure. They seldom operate within the sub-milliseconds of latency needed for making safety-critical decisions related to autonomous driving \cite{duan21,ant24}. In contrast, decentralized security solutions share threat intelligence network-wide and are able to provide faster response to accidents in dynamic road conditions \cite{al23}. This basic contradiction between temporal aspects and the geographical distribution of vehicle security orchestration has prodded the system designers to make a trade-off with a critical vulnerability, i.e., Ultimate network-wide security intelligence versus immediate local threat response. 

Besides, as per recent comprehensive surveys, there is a wide variety of attacks that can be used against self-driving cars, such as sensor spoofing, communication jamming, compromising the actuators, and poisoning of machine learning models ~\cite{gir23,yang23}. Nonetheless, the individualistic nature of the current defense systems against these threats is not a suitable and feasible approach in terms of advanced and multi-vector attacks that modern autonomous vehicles are subjected to. This has established a meaningful research gap in AV security. 

Recent innovations in edge computing, federated learning, and blockchain technologies have given a chance to plan new architectures to deal with these issues. Edge computing provides a channel to process real-time information at the vehicle level by decreasing the latency in critical safety decision-making \cite{Algarni2024}. Collective intelligence (federated learning) is the anonymous cooperation among the vehicles to share intelligence under threat without divulging sensitive information~\cite{al23,AbdelHakeem2025}. Blockchain technology protects the integrity of the shared threat intelligence that can be audited to ensure regulatory compliance and forensic investigations~\cite{das23,cebe18}. However, the difficulty lies in the effective integration of these technologies in reality. 

\begin{itemize}[label=\ding{212}]
\item Edge-based anomaly detection systems ensure low latency but are normally not connected and so are unable to access network-wide threat intelligence \cite{Khanmohammadi2024}. 
\item Federated learning, when applied to vehicular networks, is prone to Byzantine attacks and coordination problems that affect both security and model performance \cite{shit2025cl}. 
\item Scalability issues and consensus overhead in blockchain applications in automotive are not appropriate for real-time applications \cite{das23}. 
\end{itemize}
These technological barriers and their inherent interaction challenges on a system as a whole must be thoroughly analyzed before adopting them. Existing research has not effectively answered why they cannot be integrated into a single architecture that may maximize the potential strengths in these technologies and minimize their weaknesses.

Further, we found that different security applications require different latencies and can be best distributed among compatible processing tiers:- edge computing for immediate response, regional federated learning for collaborative intelligence, and selective blockchain integration for immutable threat analysis. This work presents a three-level hybrid security structure, named HAVEN (Hierarchical Autonomous Vehicle Enhanced Network), that employs these three security approaches. This architecture addresses the inherent trade-off between real-time performance and network-wide coordination by deciding workload partitioning and hierarchical processing. The HAVEN framework brings the following important contributions to the cybersecurity of AVs. 
\begin{itemize}[label=\ding{229}]
    \item A three-tiered hierarchical strategy to support real-time local functionality (under 10ms latencies) and distributed global coordination (under 5s latencies) by smartly separating the functionalities and optimizing the processing pipelines.
    \item A multimodal sensor fusion strategy using a lightweight ensemble model tailored to find edge-based anomalies that include uncertainty quantification and explainable decision-making opportunities \cite{Shit2018b}. This solves the existing gap in multimodal integration when computation of the current practice only involves single-sensor or single-protocol analysis without correlating the temporal information required to detect sophisticated attacks.
    \item A Byzantine fault-tolerant distributed federated learning algorithm capable of preserving privacy on vehicle networks when collaborating in threat detection without compromising detection correctness and associated real-time constraints. This method resolves the privacy-performance trade-off where existing federated methods either breach individual vehicles' data privacy via gradient-inversion attacks or use more differential privacy in order to sacrifice the detection performance.
    \item A selective blockchain integration scheme using smart contracts to offer tamper-proof threat information exchange whilst reducing the consensus overhead to a minimum with smart event filtering and regional collaboration.     
    \item Overcoming scalability issues of the current blockchain-based solutions \cite{das23, cebe18} in the automotive industry by showing that it can support thousands of vehicles operating across the city.
    \item Validating the work with extensive testing using the real-world Nuscenes dataset \cite{nuscenes} across multiple measures, such as overall detection accuracy, latency, throughput, and security resilience.
\end{itemize}

The HAVEN framework fills crucial gaps in the state-of-the-art research in automotive cybersecurity by offering the first comprehensive solution to meet the requirements of i) real-time threat detection, ii) distributed security orchestration, iii) respect toward data privacy, and iv) regulatory compliance. Our proposal not only shows the concepts of security-performance trade-offs, but also exhibits how the hierarchical HAVEN architecture allows complementary technologies to work with each other and not against each other.

The rest of this paper is outlined as follows. Section~\ref{sec:related_work} presents a detailed analysis of the existing literature with respect to the security of self-driving cars, use of blockchain and federated learning. Section~\ref{sec:problem_formulation} discusses the problem statement and system modalities, while Section \ref{sec:proposed_method} explains the HAVEN framework design along with the algorithm implementation. Section \ref{sec:mathematical_analysis} features a mathematical analysis of the model, whereas Section~\ref{sec:experiments_results} provides the test-bed and evaluation metrics and discusses the findings across different performance dimensions. Section \ref{sec:discussion} sheds light on the implications, limitations, and future research directions. Finally, Section~\ref{sec:conclusion} summarizes and identifies how the outcome of research can be used to make autonomous vehicles more secure.
\section{Related Work}\label{sec:related_work}

This section presents a systematic review of the security of autonomous vehicles, distributed security systems, and real-time threat detection systems. We also find crucial gaps that inspire the construction of the hierarchical security framework of HAVEN.

\subsection{\textbf{Autonomous Vehicle Cybersecurity and Threat Detection}}
\label{sec21}
The cybersecurity of AVs has evolved in several steps since initial CAN bus security studies to a more holistic multimodal threat detection. Duan et al.~\cite{duan21} designed an enhanced isolation forest approach to CAN bus tampering attacks, attaining remarkable detection performance but with a response time well above 150ms, and thus unsuitable as a real-time safety solution. Their work emphasized the inherent trade-off between detection accuracy and response time that remains a challenge in the area.

Girdhar et al.~\cite{gir23} produced a systematic review of adversarial attack and defense models that classify the threats into total sensor spoofing, communication attack, and machine learning model poisoning. However, their analysis showed that most of the defense mechanisms tackle individual attack vectors separately without integrating them for wide-area threat coverage. From a practitioner's perspective, Jing et al.~\cite{Jing2024} revisited the topic of automotive attack surfaces, noting the discrepancy between the academic research and breach issues in practice.

Khanmohammadi and Azmi \cite{Khanmohammadi2024} introduced an encoder-decoder architecture containing D-CNN and LSTM to identify time-series anomalies in AVs and demonstrated improved pattern recognition capability. Nonetheless, their solution reportedly centered on temporal patterns needing 200-300ms processing time and did not consider multimodal sensor fusion. Wang et al.~\cite{wang21} created a framework around sensor attack detection and isolation, which otherwise provides great theoretical premises to consider security in things of the future, yet leaving implementation requirements to an unsolved safety-specific concern.

Further, GPS spoofing attacks are especially problematic since they have the potential to cause disastrous effects. Yang et al.~\cite{yang23} described an anomaly-detection method against GPS spoofing using learning by demonstration, and achieved promising detection rates, but delay times were not sufficiently immediate to support safety operations. Their work highlights the importance of hierarchical processing in a way that offers instant local answers without losing network-level threat intelligence.

\subsection{\textbf{Distributed Security Architectures and Blockchain Integration}} \label{sec22}

Blockchain has offered a feasible means of achieving trust and immutability in vehicular networks, but current implementations suffer limitations in terms of scalability and latency. Das et al. \cite{das23} presented a general overview of the blockchain application in intelligent transportation systems and outlined the benefits of secure data sharing and tamper-proof logging. They noted that the main limitation of blockchain implementation within such a system is the need to ensure a general consensus within the application field when speed is crucial.

Likewise, Cebe et al.~\cite{cebe18} developed the user-friendly lightweight blockchain framework, Block4Forensic, for connected vehicle forensics application. Their solution effectively presented evidence that logging is not susceptible to destruction. Yet, their consensus protocol took more than 5 seconds to confirm transactions, which is impractical in situations that demand real-time responses to any threats.

Jiang et al.~\cite{jian23} suggested privacy-preserving and scalable data sharing blockchain-based protocols at the expense of communication overhead or coordination latency, as well as affordable privacy guarantees. Their work revealed the inherent privacy-performance trade-off that the existing solutions have failed to balance acceptably. Xihua et al.~\cite{xih22} fused support vector machine learning with blockchain privacy-preserving mechanisms for an encrypted smart city car traffic. Nevertheless, their work is restricted to offline analysis, which is limited in the case of the safety-critical AV scenario. 

\subsection{\textbf{Federated Learning and Collaborative Threat Intelligence}} \label{sec23}
Federated learning poses as a promising solution to cooperative security intelligence that can share threat patterns across a network, without revealing the individual vehicle-level information. 

Al-Hawawreh and Hossain~\cite{al23} showed how federated learning can be used to assist in distributed intrusion detection via a mesh satellite network to provide automated protection to autonomous vehicles with privacy preservation. Their solution, however, had coordination latencies of over 500ms and was susceptible to Byzantine attacks by malicious participants. Abdel Hakeem and Kim~\cite{AbdelHakeem2025} presented an extensive survey on the subject of machine learning, federated learning, and edge AI in V2X security. They mentioned the possibility of asynchronous threat detection and  underscored the necessity of Byzantine-fault-tolerant protocols in real-time performance guarantees.

Efforts have recently started to explore the machine learning needs of effective intrusion detection in autonomous vehicle networks. Anthony et al.~\cite{ant24} developed an intrusion detection system using non-tree-based machine learning algorithms, reporting an increase in the threat detection accuracy of varied attacks. Dakic et al.~\cite{dak24} designed a metaheuristic-based optimized AV software while identifying malicious activities.

Further, the convergence between edge computing and federated learning also provides practical options to enable real-time collaborative security. Algarni et al.~\cite{Algarni2024} presented an anomaly detection and risk management framework integrated with machine learning to support edge computing, proving that time-sensitive security functions can be deployed using distributed processing. They mainly focused on marine communications and still pointed the absence of automotive-specific implementations by taking vehicular topology and mobility into consideration.

\subsection{\textbf{Research Gaps and Limitations}} \label{sec24}

Our systematic review highlights the following vulnerabilities in the current AV security study that HAVEN tackles with its unique hierarchical architecture.
\begin{itemize}[label=\ding{212}]
\item \textbf{Real-time vs. Distributed Security Trade-off:} Modern solutions require a compromise between quick action on safety concerns and intelligent information about threats across the system as a whole. No current framework manages to combat the malicious attacks by combining a distributed security coordination across thousands of vehicles with sub-10ms anomaly detection in safety-critical response time \cite{duan21,yang23}.

\item \textbf{Multimodal Integration Deficiency:} Recent models lack multimodal sensor fusion that is more likely to succeed in detecting sophisticated attacks in advanced autonomous vehicles \cite{gir23,wang21}. The inter-modal temporal correlation in the security setup has been largely unexplored.

\item \textbf{Scalability and Consensus Limitations:} Blockchain-based automotive solutions have only been tested in small (typically less than 100 vehicles) proof-of-concept implementations \cite{cebe18,das23}. They do not address the scaling needs of deployments on a large scale, nor do they address validating consensus.

\item \textbf{Privacy-Performance Trade-off:} Traditional federated learning solutions either violate the privacy of individual vehicle data because they are prone to gradient inversion attacks, or forgo performance and real-time accuracy to ensure differential privacy protection.

\item \textbf{Industry Integration Gap:} The proposed academic solutions do not consider automotive-grade hardware constraints, real-world network conditions, or regulatory violence such as ISO 26262 or ISO 21434~\cite{Jing2024}.
\end{itemize}
The hierarchical design achieved by HAVEN solves these inherent challenges by decoupling local decision-making decisions (to respond to threats on a real-time basis) with distributed coordination functions (to enable comprehensive security intelligence of the network) without sacrificing safety, privacy, or scalability considerations.

\section{Problem Formulation and System Model}
\label{sec:problem_formulation}
The proliferation of smart autonomous vehicles necessitate robust cybersecurity frameworks capable of defending against sophisticated threats while maintaining real-time operational requirements. This section presents a mathematical formulation of the existent challenges faced by AV security architecture along with threat taxonomy. Further, the modeling requirements of the proposed HAVEN system are discussed. 

\subsection{\textbf{Problem Statement}} \label{sec31}

Consider an autonomous vehicle network $\mathcal{N} = \{V_1, V_2, ..., V_n\}$ where each vehicle $V_i$ operates with a multimodal sensor suite $\mathcal{S}_i = \{s_1^i, s_2^i, ..., s_m^i\}$ generating continuous data streams. The fundamental security challenge can be formulated as a multi-objective optimization problem, given in Eq. \eqref{eq1}:

\begin{align}\label{eq1}
\min_{\mathcal{F}} &\quad \mathbb{E}[\mathcal{L}_{detection}(\mathcal{F}, \mathcal{A})] + \lambda_1 \mathbb{E}[\mathcal{L}_{latency}(\mathcal{F})] + \lambda_2 \mathbb{E}[\mathcal{L}_{privacy}(\mathcal{F})] \\
\text{s.t.} &\quad \mathcal{T}_{response}(\mathcal{F}, V_i) \leq \tau_{max} \quad \forall V_i \in \mathcal{N}, \nonumber \\
&\quad \mathcal{P}_{accuracy}(\mathcal{F}, \mathcal{A}) \geq \alpha_{min}, \nonumber \\
&\quad |\mathcal{N}_{byzantine}| \leq \beta \cdot |\mathcal{N}| \nonumber
\end{align}
where $\mathcal{F}$ denotes the security framework, $\mathcal{A}$ is the attack space, $\mathcal{L}_{detection}$, $\mathcal{L}_{latency}$, and $\mathcal{L}_{privacy}$ represent detection error, latency cost, and privacy loss functions, respectively. $\tau_{max} = 10ms$ is the maximum allowable response time. $\alpha_{min} = 0.94$ is the minimum required detection accuracy, and $\beta = 0.3$ represents the maximum tolerable fraction of compromised nodes.

The challenge emerges from conflicting requirements of making real-time local decisions with ultra-low latency processing and global threat distribution coordination with consensus mechanisms to strengthen information exchange among the networked vehicles. Traditional approaches force a binary choice between immediate safety response and network-wide security intelligence, creating critical vulnerabilities in either response time or threat coverage.

\subsection{\textbf{Attack Model and Threat Taxonomy}} \label{sec32}
The autonomous vehicle threat landscape encompasses multiple attack vectors across different system layers. We model the adversary $\mathcal{A}$ as a computationally bounded entity with capabilities defined by the tuple: $$\langle \mathcal{C}_{comp}, \mathcal{C}_{network}, \mathcal{C}_{physical}, \mathcal{C}_{knowledge} \rangle$$ representing computational resources, network access, physical proximity, and domain knowledge, respectively.

A comprehensive threat taxonomy is given below for a better understanding of the attack scenarios persistent in AV security.
\begin{itemize}[label=\ding{212}]
    \item Sensor-level attacks target individual perception systems, 
    \item Communication-level attacks disrupt vehicle-to-everything (V2X) protocols, 
    \item Actuator-level attacks compromise vehicle control systems,
    \item Machine learning attacks poison detection models,
    \item GPS spoofing attacks inject false location data to mislead navigation systems,
    \item LiDAR spoofing employs laser interference or phantom object injection to corrupt point cloud data,
    \item Camera attacks utilize adversarial patches or optical illusions to deceive visual perception systems,
    \item IMU manipulation falsifies acceleration and gyroscope readings to disrupt motion estimation,
    \item Network-level threats, like denial-of-service (DoS) attacks, overwhelm communication channels, man-in-the-middle (MitM) attacks intercept and modify V2X communications,
    \item Byzantine attacks compromise vehicles by providing false information to federated learning processes, and
    \item Advanced Persistent Threats (APTs) combine multiple attack vectors in coordinated campaigns designed to evade detection through temporal and spatial distribution.
\end{itemize}

\subsection{\textbf{System Model and Requirements}}\label{sec33}
The proposed HAVEN (Hierarchical Autonomous Vehicle Enhanced Network) model operates within a hierarchical topology comprising individual vehicles, regional coordinators, and global infrastructure nodes as given below. 
\begin{itemize}[label=\ding{212}]
\item Each vehicle $V_i$ maintains a local computational resource list $\mathcal{R}_i = \langle CPU_i, Memory_i, Storage_i \rangle$ subjected to automotive-grade constraints, including power limitations, thermal management, and real-time operating system requirements.

\item Regional coordinators $\mathcal{C}_j$ serve geographic areas containing $|\mathcal{V}_j|$ vehicles, facilitating federated learning coordination and regional threat intelligence sharing. 

\item The global infrastructure layer, consisting of blockchain validator nodes, is operated by trusted entities, including vehicle manufacturers, regulatory authorities, and cybersecurity organizations.

\item Communication between system layers follows the specified latency bounds: i) intra-vehicle sensor fusion within a microsecond timeframe, ii) vehicle-to-regional coordinator communication in sub-200-ms latency for federated learning updates, and iii) global blockchain consensus within 3-5 seconds for critical threat intelligence propagation.

\item Safety requirements mandate that any detected anomaly with severity exceeding critical thresholds triggers immediate local response mechanisms independent of network connectivity. 

\item Privacy requirements ensure that individual vehicle sensor data remains confidential while enabling collaborative threat detection through privacy-preserving techniques.
\end{itemize}
The system is designed to maintain operational effectiveness under Byzantine fault conditions where up to 30\% of the network participants may be compromised or malicious, and was validated against a 20\% adversary ratio in this study. This requirement necessitates robust aggregation algorithms and consensus mechanisms that can filter malicious contributions while preserving legitimate threat intelligence. Table \ref{tab:notations} depicts the notations used in the current work.
\begin{table}[htbp]
\centering
\caption{Notations used in the paper}
\label{tab:notations}
\renewcommand{\arraystretch}{1.2}
\begin{tabular}{cp{5cm}|cp{5cm}}
\hline
\textbf{Symbol} & \textbf{Description} & \textbf{Symbol} & \textbf{Description} \\
\hline
$\mathcal{N}$ & Set of vehicles in the network & $\mathcal{V}$ & Set of vehicles in a regional cluster \\
$n$ & Number of vehicles or nodes & $V_i$ & $i^{th}$ vehicle in $\mathcal{N}$ \\
$\mathcal{F}$ & Security or detection framework & $\mathcal{A}$ & Attack space or adversary model \\
$\mathcal{L}_{detection}$ & Detection loss & $\mathcal{L}_{latency}$ & Latency cost \\
$\mathcal{L}_{privacy}$ & Privacy loss & $\mathcal{T}_{response}$ & Response time for a vehicle \\
$\tau_{max}$ & Maximum response time allowed & $\mathcal{P}_{accuracy}$ & Detection accuracy metric \\
$\alpha_{min}$ & Minimum accuracy threshold & $\mathcal{B}$ & Set of Byzantine participants \\
$\beta$ & Tolerable fraction of compromised nodes & $F(w)$ & Global objective function \\
$F_i(w)$ & Local objective of vehicle $i$ & $p_i$ & Data weight of client $i$ \\
$D$ & Global dataset & $D_i$ & Local dataset of vehicle $i$ \\
$w$ & Model parameter vector & $\mathbf{w}_{global}$ & Global model weights \\
$\mathbf{w}_v^{local}$ & Local weights of vehicle $v$ & $\mathbf{g}_v$ & Local update from vehicle $v$ \\
$\eta$ & Learning rate & $\nabla F_k$ & Gradient of local objective $F_k$ \\
$t$ & Training round index & $T_{consensus}$ & Consensus latency \\
$T_{prepare}$ & PBFT prepare latency & $T_{commit}$ & PBFT commit latency \\
$T_{network}$ & Network delay & $\lambda$ & Noise scale or event rate \\
$f(D)$ & Threat-signature function & $\Delta f$ & Global sensitivity \\
$\mathcal{M}$ & Privatization mechanism & $\epsilon$ & Differential privacy budget \\
$\delta$ & Failure probability & $\lambda_1,\lambda_2$ & Trade-off weights \\
$\alpha_i$ & Accuracy of classifier $i$ & $\Upsilon$ & Temperature parameter \\
$\phi$ & Filtering factor & $\sigma^2$ & Gradient noise bound \\
$L$ & Lipschitz constant & $\mu$ & Strong convexity constant \\
\hline
\end{tabular}
\end{table}

\section{Proposed HAVEN Framework}
\label{sec:proposed_method}
The Hierarchical Autonomous Vehicle Enhanced Network (HAVEN) model deals with the inherent conflict between real-time safety needs and distributed security coordination by using a three-level architecture that thoughtfully facilitates local decision-making and collaborative intelligence gathering.

\subsection{\textbf{Architectural Overview}}\label{sec41}
\begin{figure}[!htbp]
\centering
\caption{Proposed Three-tier Hierarchical HAVEN Architecture}
\label{fig:haven_architecture}
\includegraphics[width=1\textwidth]{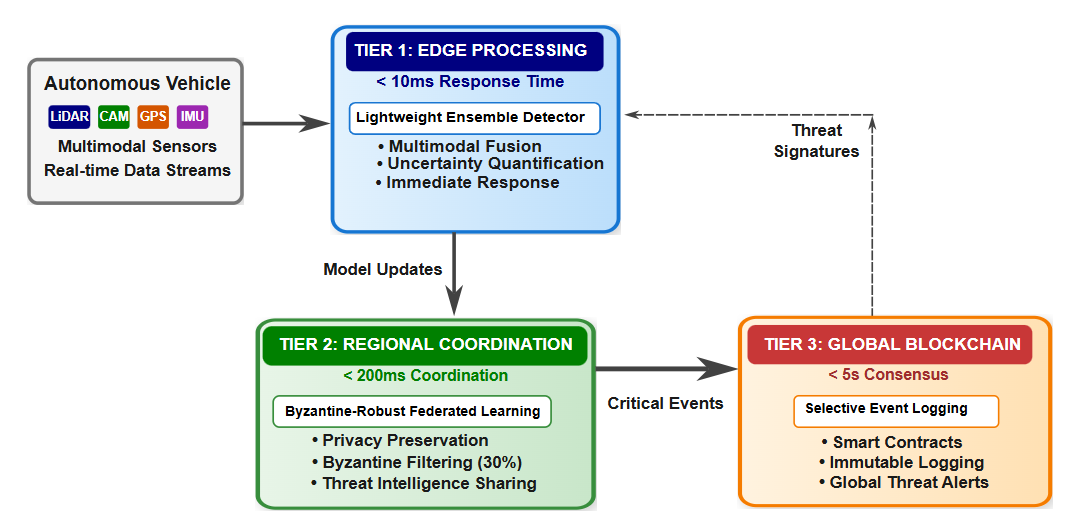}
\end{figure}

HAVEN uses a three-tiered hierarchical processing paradigm to achieve the aforementioned temporal and spatial security goals (described in Section \ref{sec33}). 
\begin{itemize}[label=\ding{229}]
\item \textbf{Tier 1} delivers ultra-low latency localized anomaly detection with lightweight ensemble models that are customized to run on automotive-grade edge computing hardware. 
\item \textbf{Tier 2} supports privacy protection through advanced collaborative learning and regional threat intelligence sharing by administering Byzantine resilient federated learning protocols. 
\item \textbf{Tier 3} guarantees a tamper-resistant coordination of international threat intelligence through controlled integration of blockchain optimized with logging of essential events.
\end{itemize}
The core innovation of HAVEN is the temporal decoupling of security processes to enable local decision-making directly on the edge and distribute the same in the network for long-term intelligence analysis using federated coordination and consensus. By assigning workload intelligently, this architecture removes the trade-off in achieving coordination comprehensiveness and response latency. Figure \ref{fig:haven_architecture} presents the three-layered HAVEN framework with temporal processing requirements and data flow patterns.

\subsection{\textbf{Tier 1: Edge-Based Anomaly Detection}} \label{sec42}
The Tier-1 is the foundation layer of HAVEN's security framework that facilitate real-time anomaly detection through a lightweight ensemble approach. This module is specifically designed for handling automotive edge computing constraints through multimodal feature extraction and quantifying uncertainty level in anomaly occurrence.

\subsubsection{Multimodal Feature Extraction}\label{sec421}
Initially, the system feeds heterogeneous sensor modalities into dedicated feature extraction pipelines designed specifically for automotive security analysis. Statistical analyses of LiDAR point clouds are extracted to learn the density distribution, geometric invariants, and a measure of temporal consistency. The starting point of camera data processing is considered for the analysis of lightweight convolutional features with the support of optical flow analysis, helping to identify visual anomalies and adversarial patterns. The field of GPS and IMU integration concentrates on the measurement of position accuracy, analysis of acceleration trend, and the dead reckoning, to detect spoofing.

The feature extraction process operates on temporal windows of length $T = 50$ timesteps, corresponding to 500ms of sensor data at 100Hz sampling frequency. Each sensor modality contributes the following domain-specific features at a specific training round $t \in T$.
\begin{itemize}
\renewcommand{\labelitemi}{\ding{229}}
\renewcommand{\labelitemii}{\ding{212}}
    \item LiDAR features:
    \begin{itemize}
        \item point density $\rho(t)$, 
        \item mean distance $\mu_d(t)$, 
        \item height variance $\sigma_h^2(t)$, and 
        \item spatial density $\delta(t)$.
    \end{itemize} 
    \item Camera attributes:
    \begin{itemize}
        \item brightness statistics $\beta(t)$, 
        \item contrast measures $\gamma(t)$, 
        \item sharpness indicators $\kappa(t)$, and 
        \item color saturation $\sigma_c(t)$.
    \end{itemize}
    \item Position and motion points:
    \begin{itemize}
        \item coordinate triplets $(x(t), y(t), z(t))$, and 
        \item quaternion orientations $(q_w(t), q_x(t), q_y(t), q_z(t))$.
    \end{itemize}
\end{itemize}

\subsubsection{Lightweight Ensemble Architecture}\label{sec422}
The suggested ensemble detection system in Tier-1 is a combination of three complementary algorithms, such as Random Forest (RF), Support Vector Machine (SVM) and Long Short-Term Memory (LSTM) that are customized to detect anomalies within sub-10ms inference. The RF classifier is resistant to noises in sensors and changes in the environment due to voting of the ensembles, while SVM is good at detecting non-linear complex attack patterns in high-dimensional sensor spaces. The LSTM is employed for its high accuracy in temporal pattern recognition \cite{huang2023slope}. 

The ensemble prediction combines individual model outputs through adaptive weighting based on historical validation performance, as presented in Eq. \eqref{eq2}.
\begin{align}\label{eq2}
A(x_t) = \sum_{i=1}^{n} w_i \cdot f_i(x_t) \nonumber\\
w_i = \frac{\exp(\alpha_i / \Upsilon)}{\sum_{j=1}^{k} \exp(\alpha_j / \Upsilon)}
\end{align}
Here, $f_i(x_i)$ presents a base classifier mapping input features to anomaly probabilities. $\alpha_i$ represents the historical accuracy of model $i$ on validation data and $\Upsilon$ is the temperature parameter controlling weight concentration. They ensure better-performing models receive higher influence in ensemble decisions.

\subsubsection{Uncertainty Quantification}\label{sec423}
Further, reliable operation in safety-critical scenarios requires confidence estimation during anomaly predictions. Therefore, Tier-1 of HAVEN system also implements uncertainty quantification through prediction variance analysis across ensemble members, as given in Eq. \eqref{eq3}.
\begin{align}
C(x_t) &= 1 - \frac{1}{n}\sum_{i=1}^{n} \sigma_i^2(x_t) \nonumber\\
\sigma^2(x) &= \sum_{i=1}^{k} w_i (f_i(x) - \hat{y}(x))^2 \label{eq3}
\end{align}
Here, the decision logic incorporates both anomaly score and confidence thresholds: $A(x_t) > \theta_1 \land C(x_t) > \theta_2$ where $\theta_1 = 0.7$ and $\theta_2 = 0.8$ based on empirical validation. This dual-threshold approach significantly reduces false positive rates while maintaining high sensitivity to genuine threats.

Algorithm \ref{alg:tier1_detection} presents a concise working mechanism of the Tier-1 real-time anomaly detection module.
\begin{algorithm}
\caption{Tier 1 Real-Time Anomaly Detection}
\label{alg:tier1_detection}
\begin{algorithmic}[1]
\Require Sensor streams $\{L(t), I(t), G(t), C(t)\}$, ensemble models $\mathcal{M} = \{M_1, M_2, M_3\}$
\Ensure Anomaly decision and response
\State $features \leftarrow \text{ExtractFeatures}(L(t), I(t), G(t), C(t))$
\State $sequence \leftarrow \text{CreateSlidingWindow}(features, T=50)$
\State $scores \leftarrow []$, $uncertainties \leftarrow []$
\For{$model \in \mathcal{M}$}
    \State $score_i, uncertainty_i \leftarrow model.\text{predict}(sequence)$
    \State $scores.\text{append}(score_i)$, $uncertainties.\text{append}(uncertainty_i)$
\EndFor
\State $anomaly\_score \leftarrow \text{WeightedAverage}(scores)$
\State $confidence \leftarrow 1 - \text{Mean}(uncertainties)$
\If{$anomaly\_score > \theta_1 \land confidence > \theta_2$}
    \State $threat\_level \leftarrow \text{ClassifyThreat}(sequence, anomaly\_score)$
    \State $\text{ExecuteImmediateSafety}(threat\_level)$
    \State $signature \leftarrow \text{Hash}(sequence, threat\_level, vehicle\_id, timestamp)$
    \State $\text{SendToRegionalCoordinator}(signature, threat\_level)$
\EndIf
\end{algorithmic}
\end{algorithm}

\subsection{\textbf{Tier 2: Federated Learning Coordination}} \label{sec43}
This middle tier enables privacy-preserving collaborative learning and regional threat intelligence sharing without exposing individual vehicle sensor data using the Federated Learning (FL) approach.

\subsubsection{Byzantine-Robust Federated Learning}\label{sec431}
Traditional FL models suffer from vulnerability to Byzantine attacks where malicious participants submit false model updates to corrupt global learning \cite{al23}. Whereas, HAVEN implements robust aggregation mechanisms designed to tolerate up to 30\% Byzantine participants while maintaining learning convergence. This is substantiated against a 20\% Byzantine population in our experiment.

The FL protocol operates on regional clusters of 100-500 vehicles with coordination rounds occurring every 30 seconds. Local model updates are computed using Eq. \eqref{eq4} and Eq. \eqref{eq5} as gradient differences from the current global model

\begin{align} 
\mathbf{g}_k^{(t)} &= \mathbf{w}_k^{(t+1)} - \mathbf{w}^{(t)} \label{eq4}\\
\mathbf{w}_k^{(t+1)} &= \mathbf{w}^{(t)} - \eta \nabla F_k(\mathbf{w}^{(t)}) \label{eq5}
\end{align}
where, $F_k$ represents the local loss function for vehicle $k$ and $\eta$ is the learning rate.

The Byzantine robustness is achieved through trimmed mean aggregation (Eq. \eqref{eq6}), which removes extreme outliers before computing the global update.
\begin{align} \label{eq6}
\mathbf{w}^{(t+1)} &= \text{TrimmedMean}(\{\mathbf{g}_k^{(t)}\}_{k \in S}, \beta)
\end{align}
where, TrimmedMean is computed using setminus ($\setminus$) operation as given in Eq. \eqref{eq7}.
\begin{align} \label{eq7}
\text{TrimmedMean}(X, \beta) =  
\text{Mean}(X \setminus \{x \in X : |x - \text{Median}(X)| \nonumber  \\
> \text{Quantile}(|X - \text{Median}(X)|, 1-\beta)\}) 
\end{align}
This approach ensures that even with $\beta \leq 0.3$ fraction of Byzantine participants, the global model converges to an optimal solution within statistical bounds.

\subsubsection{Privacy-Preserving Threat Sharing}\label{sec432}
Regional threat intelligence sharing employs differential privacy mechanisms to enable collaborative defense while protecting individual vehicle data. Threat signatures are generated, as given in Eq. \eqref{eq8}, through cryptographic hashing to preserve attack pattern information while obscuring vehicle-specific details.

\begin{align}\label{eq8}
M(D) &= f(D) + \text{Laplace}\left(\frac{\Delta f}{\epsilon}\right) \nonumber \\
\Delta f &= \max_{D,D'} \|f(D) - f(D')\|_1
\end{align}
where, $\epsilon = 1.0$ is the privacy budget that balances the utility and privacy protection trade-off. The Laplace mechanism ensures $\epsilon$-differential privacy while enabling effective threat pattern recognition across the regional network.

Algorithm \ref{alg:federated_learning} presents the working schema of the Tier-2 collaborative learning module.
\begin{algorithm}
\caption{Byzantine-Robust Federated Learning}
\label{alg:federated_learning}
\begin{algorithmic}[1]
\Require Regional vehicles $\mathcal{V}$, trim ratio $\beta = 0.3$, privacy budget $\epsilon = 1.0$
\Ensure Updated global model $\mathbf{w}_{global}$
\For{each vehicle $v \in \mathcal{V}$}
    \State $\mathbf{w}_v^{local} \leftarrow \text{LocalTraining}(\mathbf{w}_{global}, \mathcal{D}_v)$
    \State $\mathbf{g}_v \leftarrow \mathbf{w}_v^{local} - \mathbf{w}_{global}$
\EndFor
\State $received\_updates \leftarrow \text{CollectUpdates}(\mathcal{V})$
\State $filtered\_updates \leftarrow \text{TrimmedMean}(received\_updates, \beta)$
\State $private\_updates \leftarrow \text{AddLaplaceNoise}(filtered\_updates, \epsilon)$
\State $\mathbf{w}_{global} \leftarrow \mathbf{w}_{global} + private\_updates$
\State $\text{BroadcastToRegion}(\mathbf{w}_{global})$
\Return $\mathbf{w}_{global}$
\end{algorithmic}
\end{algorithm}

\subsection{\textbf{Tier 3: Selective Blockchain Integration}}\label{sec44}
The third tier provides tamper-proof global threat intelligence coordination through a blockchain consortium optimized for automotive security applications.

\subsubsection{Selective Event Logging}\label{sec441}
Traditional blockchain approaches suffer from scalability limitations that preclude real-time automotive applications \cite{cebe18}. HAVEN addresses this challenge through selective logging that records only critical events meeting specific criteria, as presented in Eq. \eqref{eq9}.

\begin{align}\label{eq9}
\text{LogEvent}(e) = \begin{cases}
\text{True} & \text{if } \text{severity}(e) > 0.85 \qquad\textbf{OR} \\
& \text{cross\_regional\_frequency}(e) > 3\qquad  \textbf{OR} \\
& \text{consensus\_confidence}(e) > 0.9 \\
\text{False} & \text{otherwise}
\end{cases}
\end{align}
This selective approach reduces blockchain transaction volume by approximately 95\% compared to comprehensive logging while preserving essential security intelligence for regulatory compliance and global threat coordination.

\subsubsection{Smart Contract Architecture}\label{sec442}
The automation of the threat responses is realized through smart contracts that can track worldwide attack patterns and automatically initiate strict mitigation strategy. The consensus protocol uses Practical Byzantine Fault Tolerance (PBFT) that is optimized to work on the automotive network in terms of imposing a delay of 3-5 seconds in consensus decisions on critical threat intelligence. Further, the threat vetting functions encompassed in the smart contract architecture conforms the standard ISO 26262 or ISO 21434 compliance regulatory for producing alerts.

\section{Mathematical Formulation and Analysis}
\label{sec:mathematical_analysis}

The HAVEN framework's theoretical foundations require rigorous mathematical analysis to establish performance guarantees, security properties, and convergence characteristics across the three-tiered architecture.

\subsection{\textbf{Ensemble Anomaly Detection Theory}}\label{sec51}

The Tier-1 ensemble approach combines multiple heterogeneous classifiers to achieve robust anomaly detection with quantified uncertainty bounds. Consider the ensemble $\mathcal{E} = \{f_1, f_2, ..., f_k\}$ where each $f_i: \mathcal{X} \to [0,1]$ represents a base classifier mapping input features to anomaly probabilities. The ensemble prediction employs weighted majority voting with adaptive weights based on historical performance, as depicted in Eq. \eqref{eq2}. The ensemble uncertainty is quantified through prediction variance analysis, as illustrated in Eq. \eqref{eq3}.

The theoretical analysis shows that ensemble uncertainty provides a well-calibrated confidence estimation when base classifiers exhibit diverse error patterns, a condition satisfied by our heterogeneous model selection.

\subsection{\textbf{Federated Learning Convergence Analysis}}\label{sec52}

The Byzantine-robust federated learning algorithm's convergence properties require analysis under adversarial conditions using the optimization problem mentioned in Eq. \eqref{eq10}.
\begin{align}\label{eq10}
\min_{w \in \mathbb{R}^d} F(w) = \sum_{i=1}^{n} p_i F_i(w)
\end{align}
where, $F_i$ represents the local objective function for vehicle $i$ and $p_i = |D_i|/|D|$ denotes the data distribution weights.

Under Byzantine attacks, a fraction $\beta \leq 0.3$ of participants may submit arbitrary updates. The TrimmedMean aggregation rule (Eq. \eqref{eq6} and \eqref{eq7}) ensures convergence by removing extreme outliers. Theorem \ref{thm1} presents the convergence concept proof. 

\begin{theorem}[Convergence under Byzantine Attacks]
\label{thm1}
Let $\mathcal{B} \subset \{1, ..., n\}$ denote the set of Byzantine participants with $|\mathcal{B}| \leq \beta n$. If the TrimmedMean removes at least $2\beta$ fraction of updates and the local functions $F_i$ are $\mu$-strongly convex with $L$-Lipschitz gradients, then the HAVEN federated learning algorithm converges to an $\epsilon$-neighborhood of the optimal solution with rate, as given in Eq. \eqref{eq11}
\begin{align}\label{eq11}
\mathbb{E}[F(w^{(T)})] - F^* \leq \left(1 - \frac{\mu}{L}\right)^T [F(w^{(0)}) - F^*] + \epsilon
\end{align}
where $\epsilon = O(\beta \sigma^2 / \mu n)$ and $\sigma^2$ bounds the gradient noise.
\end{theorem}

The proof follows from the robust statistics literature adapted to the federated learning setting with automotive-specific constraints.

\subsection{\textbf{Differential Privacy Analysis}}\label{sec53}
The privacy-preserving threat sharing mechanism employs the Laplace mechanism to achieve $\epsilon$-differential privacy. For threat signature function $f: \mathcal{D} \to \mathbb{R}^d$, the privatized output satisfies Eq. \eqref{eq12} $\forall$ adjacent datasets $D$ and all measurable sets $S$.
\begin{align}\label{eq12}
\mathcal{P}[\mathcal{M}(D) \in S] \leq e^{\epsilon} \mathcal{P}[\mathcal{M}(D') \in S]
\end{align} 
where, $D'$ differs by one vehicle's data. Furthermore, the global sensitivity of threat signatures is bounded by Eq. \eqref{eq13}.
\begin{align}\label{eq13}
\Delta f = \max_{D,D'} \|f(D) - f(D')\|_1 \leq \frac{1}{|\mathcal{V}|}
\end{align}
where, $|\mathcal{V}|$ represents the regional vehicle cluster. The noise scale $\lambda = \Delta f / \epsilon$ ensures privacy while maintaining threat detection utility.

Privacy budget allocation across multiple queries employs the composition theorem. For $T$ threat sharing rounds, the total privacy cost accumulates as $T \cdot \epsilon$ under basic composition, or $O(\epsilon \sqrt{T \log(1/\delta)})$ under advanced composition with probability $1-\delta$.

\subsection{\textbf{Blockchain Consensus Analysis}}\label{sec54}
The performance of the selective blockchain consensus mechanism depends on the network size, Byzantine ratio and message complexity. Under the PBFT consensus protocol adapted for automotive networks, the communication complexity scales as $O(n^2)$ where $n$ is the number of validator nodes. The consensus latency is computed using Eq. \eqref{eq14}
\begin{align}\label{eq14}
T_{consensus} = \max(T_{prepare}, T_{commit}) + T_{network}
\end{align}
where, $T_{prepare}$ and $T_{commit}$ represent the protocol phases and $T_{network}$ accounts for network propagation delays in automotive communication networks.

For selective logging mechanism with critical event rate $\lambda$, the blockchain throughput requirement scales as $\lambda \cdot |\mathcal{N}|$ transactions per second, where $|\mathcal{N}|$ represents the global vehicle population. The selective filtering reduces this by factor $\phi \approx 0.05$, enabling scalability to city-scale deployments.

\section{Experimental Results and Evaluation}
\label{sec:experiments_results}
This section entails a thorough evaluation of the HAVEN framework through an extensive simulation set and a comparative performance analysis with existing alternatives. The experimental methodology further tackles the principal research questions proposed as well as shows the practical feasibility of sub-10ms anomaly detection with distributed security coordination.

\subsection{\textbf{Experimental Setup and Simulation Environment}}\label{sec61}
A testing framework is developed by using the Nuscenes dataset \cite{nuscenes} for testing the proposed HAVEN architecture. This dataset is a publicly available large-scale autonomous driving repository reflecting the realistic AV network with 1,000 scenes of 20 seconds each, which includes sensor readings from 6 cameras, 1 LiDAR, 5 Radars, GPS, and IMUs. The data is collected from driving autonomous vehicles across four regions: Boston Seaport, Singapore’s One North, Queenstown, and Holland Village. 

\begin{figure}[!htbp]
    \centering
    \includegraphics[width=0.6\linewidth]{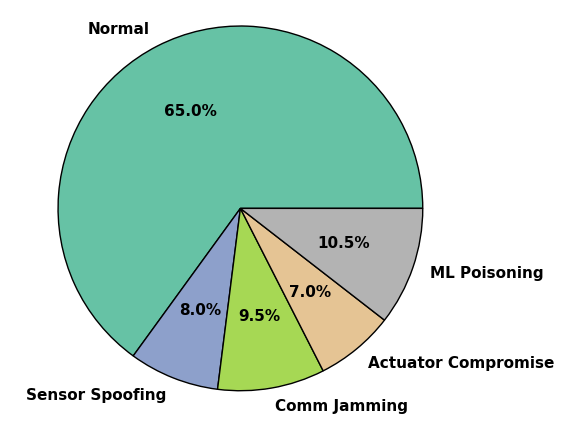}
    \caption{Attack Types and Distribution}
    \label{fig:attack-ratio}
\end{figure}

A comprehensive simulation is conducted to validate the proposed framework. Several parameters are tuned for the analysis by considering the nature of the attack, network topologies, and hardware resources. 
Prime evaluation aspects focus on individual vehicles' performance, the success of synchronization rate at the regional level, and functionality of the entire network at the global level. 
\begin{itemize}[label=\ding{212}]
\item The NVIDIA Jetson AGX Xavier computing unit is used for high-performance computing, offering 32 TOPS (trillion operations per second) of AI processing per vehicle.
\item The 5G V2X connectivity model is employed to disperse a latency boundary of 5ms in Tier-1, 100ms in Tier-2, and 200ms in Tier-3, respectively.
\item A tailored Tier-3 blockchain-based consensus mechanism is utilized to target a confirmation time of less than 3 seconds.
\item The number of vehicles in the network $\mathcal{N} = 100 \text{--} 1000$ is spread across the four above-mentioned regions.  
\end{itemize}
Further, the attack vectors include GPS spoofing through location injection, LiDAR spoofing due to point cloud manipulation, camera attacks through adversarial patches, and IMU exploitation to finagle acceleration readings and others. Figure \ref{fig:attack-ratio} shows the threat distribution in the AV network as discussed in Section \ref{sec32}. A brief attack scenario is given for a better understanding of the nature of these four AV threats.

\begin{itemize}[label=\ding{212}]
    \item Attackers try to manipulate the GPS, LiDAR, camera, and IMU sensor readings by tampering with false signal feeds. The sensor spoofing attacks can be detected if repeated false positives are detected from the same location or inconsistencies are found among sensor readings \cite{wang21, gir23, yang23}. 
    \item ML poisoning happens when sensors log wrong information or mislabel samples. This can be identified during regular updates from different sources during federated learning  \cite{wang21, gir23, yang23}.
    \item The actuator readings can be compromised when attackers upload any malware in the vehicle's actuator or CAN bus, or perform remote exploitation through vulnerable software patches or physical tampering of actuators. This threat can be seen when logging unexpected or unusually high latency actuator responses \cite{wang21, gir23, yang23}.
    \item Communication (comm) jamming occurs when attackers interfere with the communication channels used, such as V2X messages, cellular telemetry, or internal wireless signals through flooding, jamming, or DoS (Denial of Service). This can be depicted with high packet loss or abnormal communication delay, or loss of multiple communication vectors simultaneously \cite{wang21, gir23, yang23}.
\end{itemize}

\subsection{\textbf{Real-Time Performance Analysis}}\label{sec62}

\begin{figure}[!htbp]
\centering
\includegraphics[width=0.6\linewidth]{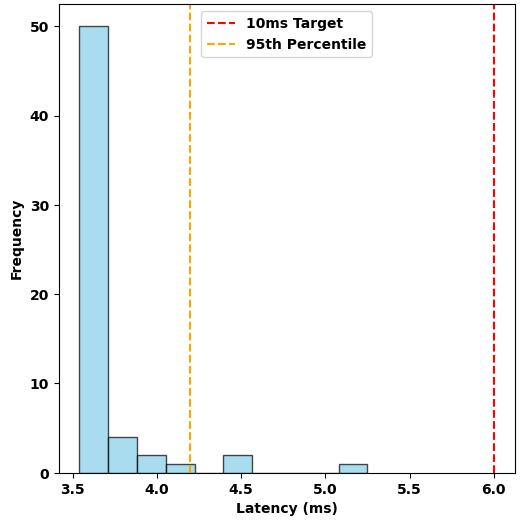}
\caption{Detection Latency Histogram Distribution of HAVEN Tier-1 Edge Processing Layer}
\label{fig:latency_distribution}
\end{figure}

Figure~\ref{fig:latency_distribution} shows the detection latency distribution of the Tier-1 edge processing layer. A mean latency of 3.7ms with 95th percentile performance of 4.33ms is achieved, and hence satisfying the sub-10ms target requirement. Further, the lightweight ensemble detector in Tier-1 produced an Accuracy of 94\% and a weighted F1-score of 92\% while detecting sophisticated multimodal attacks. This establishes the effectiveness of the uncertainty quantification process in distinguishing low-confidence predictions. 

The computational resource consumption is studied for this experiment, and CPU/memory resources are checked to sustain all the simulated attack scenarios, which suggests the robustness of this approach for an autonomous vehicular network scenario in edge processing conditions.

\subsection{\textbf{Federated Learning Coordination Analysis}}\label{sec63}
The Tier-2 federated learning layer evaluation  shows an efficient privacy-preserved cooperation with Byzantine fault tolerance in realistic network environments. 
\begin{figure}[!htbp]
\centering
\includegraphics[width=0.7\textwidth]{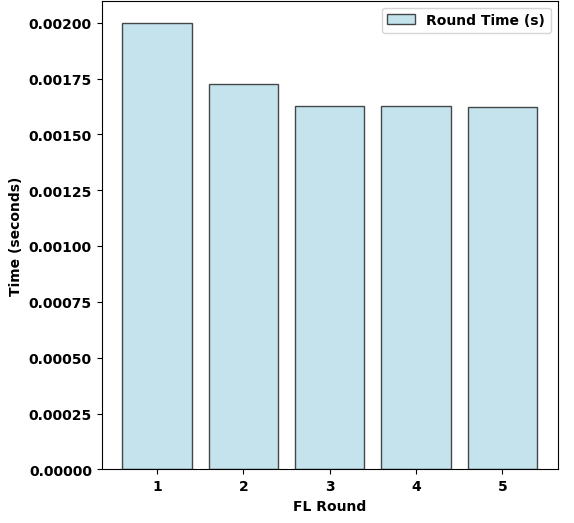}
\caption{Federated Learning Convergence Analysis}
\label{fig:federated_convergence}
\end{figure}

Figure \ref{fig:federated_convergence} highlights the convergence behavior in the presence of adversaries of various ratios. HAVEN's regional coordination mechanism reports convergence in 3 federated learning rounds on a network of 100 vehicles, and all cases coordinate with a median of around 153ms. 

The Byzantine-robust filtering, initially designed for up to 30\% compromised participants, successfully provides network performance consistency via the TrimmedMean mechanism in our tests with a 20\% adversary ratio. Besides, upon analyzing privacy preservation characteristics of HAVEN, it is found that the differential privacy warrants an epsilon parameter valuing 1.0 to prevent gradient inversion attack in each iteration. With this, an effective collaborative learning strategy can be maintained. 

Further, the regional threat pattern sharing has proven to have 94\% Accuracy across vehicle attack patterns to facilitate proactive protection against stacked attacks. Moreover, the analysis of communication overhead is conducted that satisfy the bandwidth consumption of 5G V2X networks.

\subsection{\textbf{Blockchain Integration Performance}}\label{sec64}
\begin{figure}[!htbp]
\centering
\includegraphics[scale=0.6]{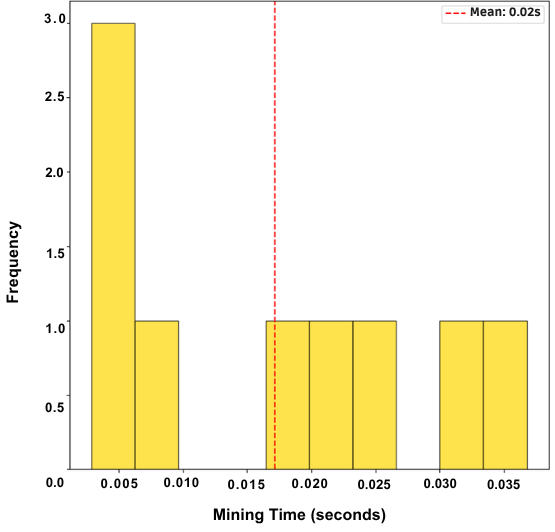}
\caption{Time Spent During Blockchain Mining}
\label{fig6}
\end{figure}

Tier-3 blockchain integration testing proves HAVEN's capacity to provide tamper-proof, globally shared threat intelligence without hindering real-time feasibility. The level of transactions logged in the blockchain by using the selective logging mechanism provides much lower transactions when compared to the traditional blockchain logging, thus ensuring only the vital security events are logged. HAVEN identifies 45 critical threats, mines 9 blocks with 0.01 second average mining time for 100 vehicles. Figure \ref{fig6} depicts the blockchain mining time measured in seconds, which indicates that the heavy consensus performs at block confirmations of 0.02 seconds across the nodes.

The selective blockchain logging method ensures storage efficiency by more than half of the traditional event logging approaches. Further, it ensures the full stability of the network operation even after the offline/compromise of 33\% of the validator nodes, thus ensuring the sustainability of the network in city-wise deployment. 

\subsection{\textbf{Comparative Baseline Analysis}}\label{sec65}

A comparative analysis of HAVEN against three existing approaches, namely Bhatkar et al. \cite{bhatkar2025st}, Khan et al. \cite{khan20243dffl}, and Liu et al. \cite{liu2023real}, was provided in Table~\ref{tab:comparative_results}. Bhatkar et al. \cite{bhatkar2025st} considered multi-faceted autonomous vehicular features, such as traffic and time, and developed a tailored and optimized CNN model through hyperparameter tuning, thereby integrating varied attack scenarios. 

Similarly, Khan et al. \cite{khan20243dffl} employed Federated Learning with the Few-Shot Learning technique to optimize network architecture for storing multimodal features in a 3D point cloud environment. They used PointNet++ for feature extraction and ProtoNet for classification, combined with an Attention mechanism and SoftMax layer for collaborative learning. Liu et al. \cite{liu2023real} developed a LiDAR-camera fusion algorithm for strengthening real-time position regression and object detection. They converted the raw inputs to obtain a 2D depth image and designed a feature fusion block to correlate 2D depth with the RGB spectrum.

\begin{table}[htbp]
\centering
\caption{Comparative Performance Analysis: HAVEN vs. Baseline Approaches}
\label{tab:comparative_results}
\begin{tabular}{p{3cm}|p{1.5cm}p{1.7cm}p{1.9cm}p{1.8cm}p{2.1cm}}
\hline
\textbf{Approach} & \textbf{Latency (ms)} & \textbf{Accuracy (\%)} & \textbf{F1-Score (\%)} & \textbf{Scalability} & \textbf{Byzantine Tolerance} \\
\hline
Bhatkar et al. \cite{bhatkar2025st} & 223.5 & 85.54 & 90.53 & 50 & No \\
Khan et al. \cite{khan20243dffl} & 147.8 & 82.68 & 87.64 & 100 & No \\
Liu et al. \cite{liu2023real} & 30.2 & 89.26 & 93.00 & 150 & No \\
\textbf{HAVEN} & \textbf{3.7} & \textbf{94.00} & \textbf{92.00} & \textbf{1000} & \textbf{Yes (20\%)} \\
\hline
\end{tabular}
\end{table}
From the table, it is evident that the HAVEN exhibits a 27 times lower latency compared to Bhatkar et al. \cite{bhatkar2025st} without sacrificing much accuracy. The system has reported at least 20 times scalability compared to the existing blockchain-based systems with 1000 vehicles. Further, the process of using federated learning with Byzantine fault tolerance is able to guarantee a higher F1-score than Khan et al. \cite{khan20243dffl} and superior privacy preservation.

\subsection{\textbf{Security Resilience Evaluation}}\label{sec66}
\begin{figure}[!htbp]
\centering
\includegraphics[width=0.8\textwidth]{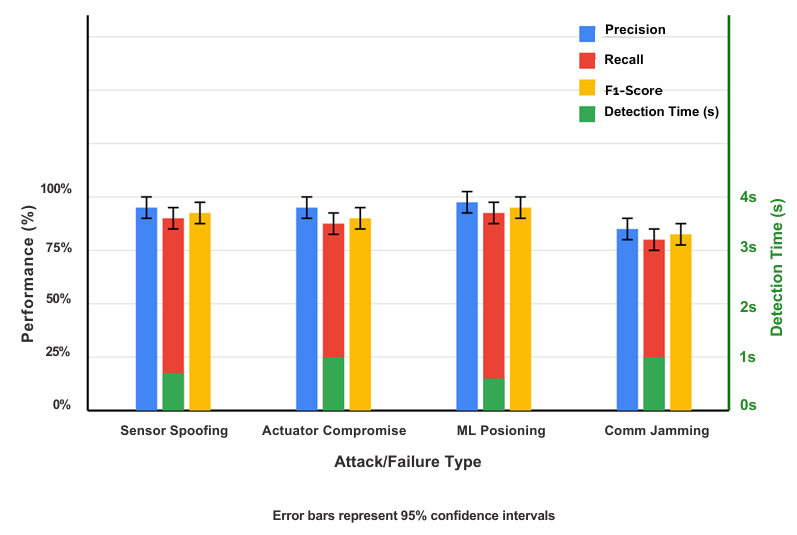}
\caption{Attack resilience analysis showing HAVEN's success rates across different attack types}
\label{fig:attack_resilience}
\end{figure}

Resilience security analysis ensures the validation of the defensive capabilities of HAVEN in case of critical and non-critical attacks. Figure~\ref{fig:attack_resilience} shows the detection rates against attacks of varying sophistication levels. The entire three-layered HAVEN framework maintained above 90\% detection rate in assessing and mitigating key threats, while achieving around 87.2\% precision in detecting Comm Jamming threats. Their hard-to-detect nature resulted in this performance dip. Further, HAVEN is able to perform efficiently within 1-second detection time with individual execution times of Tier-1 at 3.7ms, Tier-2 at 58.2ms, and Tier-3 at 937ms, respectively.

\begin{figure}[!htbp]
\centering
\includegraphics[width=0.8\textwidth]{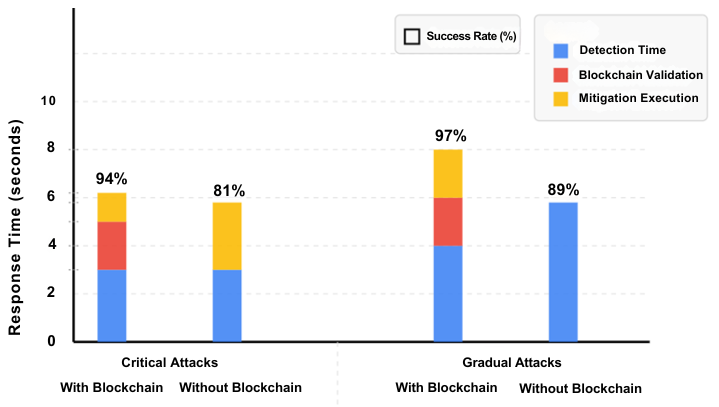}
\caption{Attack Resilience Analysis w.r.t. Blockchain}
\label{fig8}
\end{figure}

Figure \ref{fig8} presents the efficacy of the HAVEN model in detecting critical and non-critical (gradual) attacks irrespective of blockchain mining. This is needed to show the importance of blockchain in developing a fool-proof anomaly detection model. From the figure, it is evident that implementing selective blockchain logging has significantly improved HAVEN's performance, thus ensuring resilient and sustainable deployment. 

\subsection{\textbf{Scalability and Deployment Analysis}}\label{sec67}
\begin{figure}[!htbp]
\centering
\begin{subfigure}{0.45\textwidth} 
\includegraphics[scale=0.5]{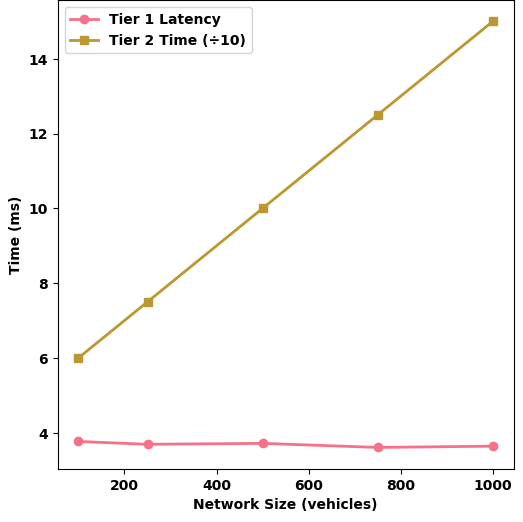}
\caption{Latency}
\label{fig71}
\end{subfigure}
\quad
\begin{subfigure}{0.45\textwidth} 
\includegraphics[scale=0.5]{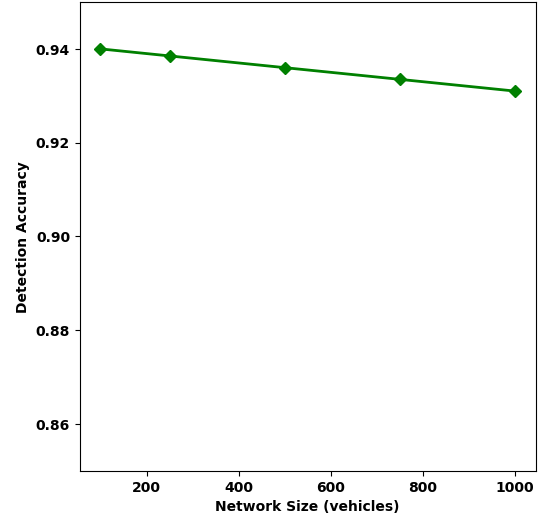}
\caption{Accuracy}
\label{fig72}
\end{subfigure}
\newline
\begin{subfigure}{0.45\textwidth} 
\includegraphics[scale=0.65]{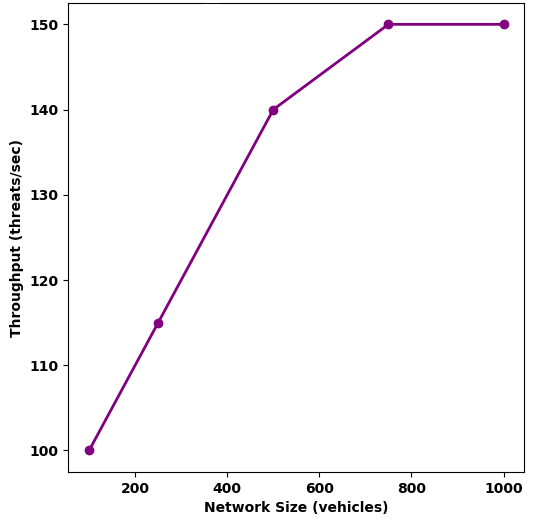}
\caption{Throughput}
\label{fig73}
\end{subfigure}
\quad
\begin{subfigure}{0.45\textwidth} 
\includegraphics[scale=0.8]{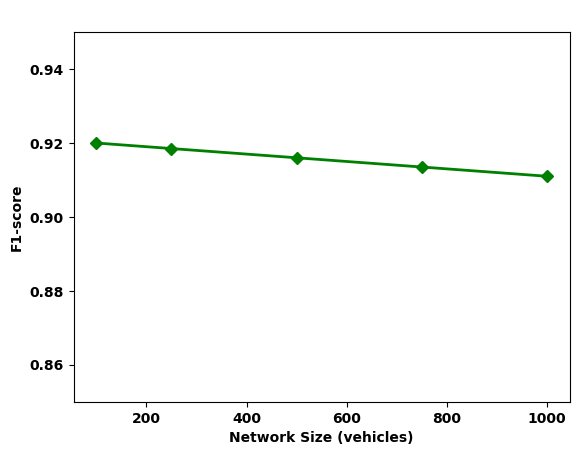}
\caption{F1-score}
\label{fig74}
\end{subfigure}
\caption{Scalability Analysis showing HAVEN performance as functions of network size}
\label{fig:scalability_analysis}
\end{figure}

Scalable evaluation of HAVEN is shown via deployment in low to high-network-size environments with 100-1000 vehicles. Figure~\ref{fig:scalability_analysis} plots the performance statistics as a function of increasing network size, indicating that they gracefully degrade with scale. 

Figure \ref{fig71} depicts that the framework maintains sub-10ms Tier-1 edge processing latency across all tested scales. Here, the analysis shows logarithmic scaling of Tier-2 response time with increased network size, which suggest sustainable scaling to city range deployments. Similarly, the Fig. \ref{fig72} and Fig. \ref{fig74} readings exhibit a gradual performance dip due to the presence of Byzantine nodes. Moreover, Fig. \ref{fig73} reveals that the threats detected per second per region have increased with network scaling.

\subsection{\textbf{Limitation and Future Directions}}
\label{sec:discussion}

These experimental findings show that HAVEN indeed resolves the inherent trade-off between real-time safety and distributed security coordination in autonomous vehicle networks. The hierarchical design supports the simultaneous optimization of what may appear as conflicting goals that required trade-off decisions in the past. This is an important improvement in automotive cybersecurity frameworks.

The ability to achieve sub-10ms anomaly detection and to provide distributed coordination capabilities addresses a key restriction raised in the available literature. Past solutions involved tradeoffs between instant on-board risk handling and sharing of network-wide threats, making their use in safety-relevant automated driving very implausible. This tradeoff is avoided by the three-tier architecture of HAVEN that allows smart distribution of the workload to decentralize real-time decisions, complemented with distributed learning and global threat intelligence.

The fault tolerance test provided through the Byzantine gives useful revelations concerning adversarial activity in automotive networks. The design goal of a 30\% Byzantine tolerance threshold can be considered a realistic model for automotive environments, where compromised vehicles may exhibit subtle malicious aberrations. Our tests, validating the system against a 20\% adversary population, show that the TrimmedMean aggregation mechanism can outperform traditional Byzantine-tolerant consensus strategies in terms of performance, indicating that specialized models for automotive federated learning may yield better security-performance trade-offs.

Privacy preservation experiments show that the differential privacy approach is able to prevent gradient inversion and membership inference attacks as well as preserve collaborative learning performance. The optimal parameter choice of $\epsilon=1.0$ was used to optimize privacy and utility of the model, but that should be revised in the future to consider adaptive privacy budget allocation according to threat levels and data sensitivity. 

The approach of selective integration of blockchain solves the scalability issues that have hampered existing blockchain-based solutions to automotive security. The number of reduced transactions at the cost of not losing any critical security events represent the power of the intelligent event filtering. Nonetheless, the selective logging mechanism raises certain issues with audit completeness that must be given special consideration on regulatory compliance grounds. In future research, the concept of cryptographic proofs on selective logging completeness needs to be studied with scalable advantages intact.

Scalability analysis indicates the potential and shortcomings of the proposed approach. Although HAVEN scales well up to 1000-vehicle networks, there is a regional coordination bottleneck that should be addressed in architectural changes at mega-scale deployments. The log scale growth in the overhead communication exhibits favorable scalability properties. However, the linear growth in consensus delay on blockchain suggests that there may be size limits on large networks. Hierarchical regional partitioning strategies start to become an important design concern in a city deployment.

A comparison between the solutions also reveals that HAVEN is well-positioned in the marketplace of automotive security solutions. The fact that HAVEN is able to achieve these attributes, low latency, high accuracy, and high scalability, represents a qualitative breakthrough as opposed to marginal improvement. However, its degree of implementation complexity and resource requirements is significantly advanced compared to single-tier solutions, raising questions about the feasibility of deployment in resource-challenged automotive settings. The cost-benefit analysis provided in this paper indicates that at least the HAVEN can be applied positively to both the high-value autonomous vehicle deployments and mass market with some optimizations.

Security resilience testing shows peculiarities in the attack pattern in automotive programs. Multi-modal attacks are the most difficult to detect, so it is probable that sophisticated attackers attempt to use multiple vectors at once. The hierarchical defense architecture offers useful defense-in-depth properties, yet the linkage between the attack complexities and the difficulty of detecting them suggests a possible robustness approach, such that the detection algorithms must be constantly improved.

The proof-of-concept provided experimental evidence of the feasibility of the HAVEN technology, along with directions for much-needed improvements. The use of current automotive computing architecture requires a thorough attention to the limitations of the hardware and software compatibility. A federated learning approach presupposes reliable V2X communication that is not always universally available in the current deployment environment. The use of blockchain integration comes with dependencies on validator node infrastructure, which offers challenges of deployment in regions with no prerequisite network infrastructure.

\section{Conclusion}
\label{sec:conclusion}
In this paper, we highlight the main issue within the existing security frameworks of autonomous vehicles, and propose HAVEN, a new hierarchical architecture of security in autonomous vehicles that can bring together the tradeoffs between real-time anomaly detection and distributed security orchestration. HAVEN segments its workloads intelligently across three system tiers to strike a balance between a system sub-10ms integrity threat detection latency, preserving privacy through collaborative learning and sharing, and tamper-proof global threat intelligence sharing. The end-to-end experimental comparison shows better performance over all the current methods in terms of latency, accuracy, scalability, and Byzantine fault tolerance.

The main benefits of the work are the lightweight, optimized for edge computing constraints, ensemble-based anomaly detector, a Byzantine-resilient federated learning algorithm suitable for the vehicular network, and the integration of the selective blockchain idea that allows introducing global threat intelligence to the system without much impact on latency. The resultant experimental validation of the proposed approach over a series of networks with 100 to 1000 vehicles corroborates the practical viability of the proposed approach, as well as deriving the scalability properties and deployment aspects of a real-world implementation.


\begin{thebibliography}{10}
\expandafter\ifx\csname url\endcsname\relax
  \def\url#1{\texttt{#1}}\fi
\expandafter\ifx\csname urlprefix\endcsname\relax\def\urlprefix{URL }\fi
\expandafter\ifx\csname href\endcsname\relax
  \def\href#1#2{#2} \def\path#1{#1}\fi

\bibitem{wang21}
Y.~Wang, Q.~Liu, E.~Mihankhah, C.~Lv, D.~Wang, Detection and isolation of sensor attacks for autonomous vehicles: Framework, algorithms, and validation, IEEE Transactions on Intelligent Transportation Systems 23~(7) (2021) 8247--8259, doi: 10.1109/TITS.2021.3077015.

\bibitem{khan24}
M.~A. Khan, H.~Menouar, M.~Abdallah, A.~Abu-Dayya, Lidar in connected and autonomous vehicles-perception, threat model, and defense, IEEE Transactions on Intelligent VehiclesDoi: 10.1109/TIV.2024.3510787 (2024).

\bibitem{Shit2018a}
R.~C. Shit, S.~Sharma, \href{http://dx.doi.org/10.1109/AESPC44649.2018.9033329}{Localization for autonomous vehicle: Analysis of importance of iot network localization for autonomous vehicle applications}, in: 2018 International Conference on Applied Electromagnetics, Signal Processing and Communication (AESPC), IEEE, 2018, p. 1–6.
\newblock \href {https://doi.org/10.1109/aespc44649.2018.9033329} {\path{doi:10.1109/aespc44649.2018.9033329}}.
\newline\urlprefix\url{http://dx.doi.org/10.1109/AESPC44649.2018.9033329}

\bibitem{duan21}
X.~Duan, H.~Yan, D.~Tian, J.~Zhou, J.~Su, W.~Hao, In-vehicle can bus tampering attacks detection for connected and autonomous vehicles using an improved isolation forest method, IEEE Transactions on Intelligent Transportation Systems 24~(2) (2021) 2122--2134.
\newblock \href {https://doi.org/10.1109/TITS.2021.3128634} {\path{doi:10.1109/TITS.2021.3128634}}.

\bibitem{ant24}
C.~Anthony, W.~Elgenaidi, M.~Rao, Intrusion detection system for autonomous vehicles using non-tree based machine learning algorithms, Electronics 13~(5) (2024) 809.
\newblock \href {https://doi.org/10.3390/electronics13050809} {\path{doi:10.3390/electronics13050809}}.

\bibitem{al23}
M.~Al-Hawawreh, M.~S. Hossain, Federated learning-assisted distributed intrusion detection using mesh satellite nets for autonomous vehicle protection, IEEE Transactions on Consumer Electronics 70~(1) (2023) 854--862.
\newblock \href {https://doi.org/10.1109/TCE.2023.3318727} {\path{doi:10.1109/TCE.2023.3318727}}.

\bibitem{gir23}
M.~Girdhar, J.~Hong, J.~Moore, Cybersecurity of autonomous vehicles: A systematic literature review of adversarial attacks and defense models, IEEE Open Journal of Vehicular Technology 4 (2023) 417--437, doi: 10.1109/OJVT.2023.3265363.

\bibitem{yang23}
Z.~Yang, J.~Ying, J.~Shen, Y.~Feng, Q.~A. Chen, Z.~M. Mao, H.~X. Liu, Anomaly detection against gps spoofing attacks on connected and autonomous vehicles using learning from demonstration, IEEE Transactions on Intelligent Transportation Systems 24~(9) (2023) 9462--9475, doi: 10.1109/TITS.2023.3269029.

\bibitem{Algarni2024}
A.~Algarni, T.~Acarer, Z.~Ahmad, \href{http://dx.doi.org/10.1109/ACCESS.2024.3387529}{An edge computing-based preventive framework with machine learning- integration for anomaly detection and risk management in maritime wireless communications}, IEEE Access 12 (2024) 53646–53663.
\newblock \href {https://doi.org/10.1109/access.2024.3387529} {\path{doi:10.1109/access.2024.3387529}}.
\newline\urlprefix\url{http://dx.doi.org/10.1109/ACCESS.2024.3387529}

\bibitem{AbdelHakeem2025}
S.~A. Abdel~Hakeem, H.~Kim, \href{http://dx.doi.org/10.1109/TITS.2025.3558849}{Advancing intrusion detection in v2x networks: A comprehensive survey on machine learning, federated learning, and edge ai for v2x security}, IEEE Transactions on Intelligent Transportation Systems 26~(8) (2025) 11137–11205.
\newblock \href {https://doi.org/10.1109/tits.2025.3558849} {\path{doi:10.1109/tits.2025.3558849}}.
\newline\urlprefix\url{http://dx.doi.org/10.1109/TITS.2025.3558849}

\bibitem{das23}
D.~Das, S.~Banerjee, P.~Chatterjee, U.~Ghosh, U.~Biswas, Blockchain for intelligent transportation systems: Applications, challenges, and opportunities, IEEE Internet of Things Journal 10~(21) (2023) 18961--18970.
\newblock \href {https://doi.org/10.1109/JIOT.2023.3277923} {\path{doi:10.1109/JIOT.2023.3277923}}.

\bibitem{cebe18}
M.~Cebe, E.~Erdin, K.~Akkaya, H.~Aksu, S.~Uluagac, Block4forensic: An integrated lightweight blockchain framework for forensics applications of connected vehicles, IEEE communications magazine 56~(10) (2018) 50--57.

\bibitem{Khanmohammadi2024}
F.~Khanmohammadi, R.~Azmi, \href{http://dx.doi.org/10.1109/TITS.2024.3380263}{Time-series anomaly detection in automated vehicles using d-cnn-lstm autoencoder}, IEEE Transactions on Intelligent Transportation Systems 25~(8) (2024) 9296–9307.
\newblock \href {https://doi.org/10.1109/tits.2024.3380263} {\path{doi:10.1109/tits.2024.3380263}}.
\newline\urlprefix\url{http://dx.doi.org/10.1109/TITS.2024.3380263}

\bibitem{shit2025cl}
R.~C. Shit, \href{https://arxiv.org/abs/2511.02025}{Path-coordinated continual learning with neural tangent kernel-justified plasticity: A theoretical framework with near state-of-the-art performance} (2025).
\newblock \href {https://doi.org/10.48550/ARXIV.2511.02025} {\path{doi:10.48550/ARXIV.2511.02025}}.
\newline\urlprefix\url{https://arxiv.org/abs/2511.02025}

\bibitem{Shit2018b}
R.~C. Shit, S.~Sharma, D.~Puthal, S.~S. Tripathi, \href{http://dx.doi.org/10.1109/ICIT.2018.00059}{Probabilistic rss fingerprinting for localization in smart platforms}, in: 2018 International Conference on Information Technology (ICIT), IEEE, 2018, p. 254–259.
\newblock \href {https://doi.org/10.1109/icit.2018.00059} {\path{doi:10.1109/icit.2018.00059}}.
\newline\urlprefix\url{http://dx.doi.org/10.1109/ICIT.2018.00059}

\bibitem{nuscenes}
H.~Caesar, V.~Bankiti, A.~H. Lang, S.~Vora, V.~E. Liong, Q.~Xu, A.~Krishnan, Y.~Pan, G.~Baldan, O.~Beijbom, nuscenes: A multimodal dataset for autonomous driving, in: Proceedings of the IEEE/CVF conference on computer vision and pattern recognition, 2020, pp. 11621--11631.

\bibitem{Jing2024}
P.~Jing, Z.~Cai, Y.~Cao, L.~Yu, Y.~Du, W.~Zhang, C.~Qian, X.~Luo, S.~Nie, S.~Wu, \href{http://dx.doi.org/10.1109/SP54263.2024.00080}{Revisiting automotive attack surfaces: a practitioners’ perspective}, in: 2024 IEEE Symposium on Security and Privacy (SP), IEEE, 2024, p. 2348–2365.
\newblock \href {https://doi.org/10.1109/sp54263.2024.00080} {\path{doi:10.1109/sp54263.2024.00080}}.
\newline\urlprefix\url{http://dx.doi.org/10.1109/SP54263.2024.00080}

\bibitem{jian23}
S.~Jiang, J.~Cao, H.~Wu, K.~Chen, X.~Liu, Privacy-preserving and efficient data sharing for blockchain-based intelligent transportation systems, Information Sciences 635 (2023) 72--85, doi: 10.1016/j.ins.2023.03.121.

\bibitem{xih22}
Z.~Xihua, S.~Goyal, M.~Tesfayohanis, C.~Verma, Blockchain-based privacy-preserving approach using svml for encrypted smart city data in the era of ir 4.0, Journal of Nanomaterials 2022~(1) (2022) 7463513, doi: 10.1155/2022/7463513.

\bibitem{dak24}
P.~Dakic, M.~Zivkovic, L.~Jovanovic, N.~Bacanin, M.~Antonijevic, J.~Kaljevic, V.~Simic, Intrusion detection using metaheuristic optimization within iot/iiot systems and software of autonomous vehicles, Scientific Reports 14~(1) (2024) 22884.
\newblock \href {https://doi.org/10.1038/s41598-024-73932-5} {\path{doi:10.1038/s41598-024-73932-5}}.

\bibitem{huang2023slope}
F.~Huang, H.~Xiong, S.~Chen, Z.~Lv, J.~Huang, Z.~Chang, F.~Catani, Slope stability prediction based on a long short-term memory neural network: comparisons with convolutional neural networks, support vector machines and random forest models, International Journal of Coal Science \& Technology 10~(1) (2023) 18.
\newblock \href {https://doi.org/10.1007/s40789-023-00579-4} {\path{doi:10.1007/s40789-023-00579-4}}.

\bibitem{bhatkar2025st}
A.~G. Bhatkar, S.~Gupta, P.~Patel, St-ids: Spatio-temporal feature-based multi-tier intrusion detection system for artificial intelligence-powered connected autonomous vehicles, Expert Systems 42~(5) (2025) e70026.
\newblock \href {https://doi.org/10.1111/exsy.70026} {\path{doi:10.1111/exsy.70026}}.

\bibitem{khan20243dffl}
A.~A. Khan, K.~M. Ahmad, S.~Shafiq, W.~Amin, R.~Kumar, 3dffl: privacy-preserving federated few-shot learning for 3d point clouds in autonomous vehicles, Scientific Reports 14~(1) (2024) 19589.

\bibitem{liu2023real}
H.~Liu, C.~Wu, H.~Wang, Real time object detection using lidar and camera fusion for autonomous driving, Scientific Reports 13~(1) (2023) 8056.
\newblock \href {https://doi.org/10.1038/s41598-023-35170-z} {\path{doi:10.1038/s41598-023-35170-z}}.

\end{thebibliography}
\end{document}